\documentclass[twocolumn,superscriptaddress,nofootinbib,amsmath,amssymb]{revtex4-1}

\usepackage{euscript,amssymb,amsmath}
\usepackage{amsfonts}
\newcommand{\diag}{\mbox{\rm diag}\,}

\usepackage{graphicx}
\usepackage{epsfig}

\usepackage{hyperref}
\usepackage{color}
\usepackage[normalem]{ulem}

\newcommand{\be}[1]{\begin{equation}\label{#1}}
\newcommand{\ee}{\end{equation}}
\newcommand{\ba}[1]{\begin{eqnarray}\label{#1}}
\newcommand{\ea}{\end{eqnarray}}
\newcommand{\rf}[1]{(\ref{#1})}
\newcommand{\nn}{\nonumber}

\newcommand{\de}{\partial}

\newcommand{\om}{\omega}

\begin{document}

\title{Weak field limit of higher dimensional massive Brans-Dicke gravity:\\ Observational constraints}

\author{\"{O}zg\"{u}r Akarsu}
\email{akarsuo@itu.edu.tr}
\affiliation{Department of Physics, Istanbul Technical University, Maslak 34469 Istanbul, Turkey}

\author{Alexey Chopovsky}
\email{a.chopovsky@yandex.ru}
\affiliation{Astronomical Observatory, Odessa National University, 2 Dvoryanskaya Street, Odessa 65082, Ukraine}

\author{Valerii Shulga}
\email{shulga@rian.kharkov.ua}
\affiliation{International Center of Future Science of the Jilin University, 2699 Qianjin Street, 130012, Changchun City, China}
\affiliation{Institute of Radio Astronomy of National Academy of Sciences of Ukraine, 4 Mystetstv Street, 61002 Kharkiv, Ukraine }

\author{Ezgi Yal\c{c}{\i}nkaya}
\email{ezgicanay@itu.edu.tr}
\affiliation{Department of Physics, Istanbul Technical University, Maslak 34469 Istanbul, Turkey}

\author{Alexander Zhuk}
\email{ai.zhuk2@gmail.com}
\affiliation{Astronomical Observatory, Odessa
National University, 2 Dvoryanskaya Street, Odessa 65082, Ukraine}
\affiliation{Department of Physics, Istanbul Technical University, Maslak 34469 Istanbul, Turkey}

\begin{abstract}
We consider higher-dimensional massive Brans-Dicke theory with Ricci-flat internal space. The background model is perturbed by a massive gravitating source which is pressureless in the external (our space) but has an arbitrary equation-of-state parameter $\Omega$ in the internal space. We obtain the exact solution of the system of linearized equations for the perturbations of the metric coefficients and scalar field. For a massless scalar field, relying on the fine-tuning between the Brans-Dicke parameter $\omega$ and $\Omega$, we demonstrate that (i) the model does not contradict gravitational tests relevant to the parameterized post-Newtonian parameter $\gamma$, and (ii) the scalar field is not ghost in the case of nonzero $|\Omega|\sim O(1)$ along with the natural value $|\omega|\sim O(1)$. In the general case of a massive scalar field, the metric coefficients acquire the Yukawa correction terms, where the Yukawa mass scale $m$ is defined by the mass of the scalar field. For the natural value $\omega\sim O(1)$, the inverse-square-law experiments impose the following restriction on the lower bound of the mass: $m\gtrsim 10^{-11}\,$GeV. The experimental constraints on $\gamma$ requires that $\Omega$ must be extremely close to $-1/2$.
\end{abstract}


\maketitle


\section{\label{sec:1}Introduction}

The idea of higher-dimensional spacetime dates its history back to the pioneering works by Nordstr\"{o}m \cite{Nordstrom:1988fi}, Kaluza \cite{Kaluza:1921}, and Klein \cite{Klein:1926} arguing that the extra dimensions are unobservable since they are compact and of small length scales. This compactification approach is actively used in modern theoretical physics. In particular, the Kaluza-Klein (KK) models, based on the existence of higher dimensions, form a cornerstone of string theory and M theory \cite{Becker:2007}. They are also employed in attempts to resolve challenging problems such as the hierarchy of the fundamental interactions \cite{Arkani:1998}, the nature of dark matter \cite{Cheng:2002}, and the nature of the cosmological constant \cite{Guenther:2003}.

Obviously, any viable physical theory should be in agreement with the existing empirical data. Since the KK models are essentially the modified gravity models, they must satisfy the gravitational tests successfully passed by the usual general relativity (GR) assuming four-dimensional spacetime---e.g., the deflection of light, the Shapiro time delay, and the perihelion precession of Mercury. This aspect was investigated in a series of papers \cite{EZ:2010, CEZ:2013, CEZ:2014}, where the weak gravitational fields in KK models with compact Ricci-flat internal spaces (extra dimensions) are considered. In these works, the authors consider the post-Newtonian gravitational field created by pointlike, nonrelativistic massive sources simulating compact astrophysical objects (e.g., stars) and assume that these sources yield zero pressure in the external (noncompact) space. This is indeed a rather natural assumption, since the pressure inside the nonrelativistic astrophysical objects is much less than the corresponding energy density. Therefore, in GR, while calculating the parameterized post-Newtonian (PPN) parameters $\gamma$ and $\beta$ for orbiting test masses in the Solar System, it is sufficient to neglect the Sun's pressure \footnote{In general, matter is not assumed to be pressureless in the PPN formalism; see, e.g., Ref. \cite{Will}.} \cite{Will,Landau}. In line with that, the calculated PPN parameters $\gamma$ and $\beta$ are in very good agreement with the experimental data \cite{Will}. In particular, the PPN parameter $\gamma=1$ is in very good agreement with the  precision Shapiro time-delay experiment: $\gamma = 1+(2.1\pm2.3)\times 10^{-5}$ \cite{Bertotti:2003, Shapiro:2004, Will:2005}. Therefore, in KK models, it would also be natural to assume that the gravitating mass remains pressureless in the internal space. However, the calculations for such a model show that the PPN parameter $\gamma$ is quite different from unity: $|\gamma -1| \sim O(1)$ \cite{EZ:2010}. Then, since the equation of state (EoS) in the internal space is not known, for the sake of generality, it is assumed that some nonzero parameter $\Omega$ of the EoS applies in the internal space. For this setting of the problem, it turns out that in the KK models with Ricci-flat internal spaces, in order for $\gamma$ to have a value close to $1$, the EoS parameter $\Omega$ must be very close to $-1/2$ \cite{CEZ:2013,CEZ:2014}. To restore the value $\gamma = 1$, as in GR, it is necessary to choose $\Omega = - 1/2$, which corresponds to black strings or branes \cite{Trashen:2004,Townsend:2001,Harmark:2004,EZ:2011}. However, so far, there is no satisfactory explanation of the possible nature of such a relativistic EoS in the internal space with the parameter $\Omega$ being essentially nonzero and negative. Then, a natural question arises: Is the value $\Omega = -1/2$ inevitable for viable models (satisfying the gravitational tests) with Ricci-flat internal spaces?

To answer this question, in the present paper we modify the action of the gravitational sector. We switch from the higher-dimensional version of the Einstein-Hilbert action, considered in the previous works \cite{EZ:2010,CEZ:2013,CEZ:2014}, to a scalar-tensor model, wherein the gravity has an extra scalar degree of freedom (d.o.f.) $\Phi$ coupled to the scalar curvature $R$ nonminimally. Such models arise naturally in the context of string theory and play a significant role in present-day cosmology studies (see Refs. \cite{Maeda,Skordis} and references therein). We focus on a higher-dimensional generalization of the well-known Brans-Dicke (BD) theory \cite{BD:1961}---characterized by the BD parameter $\omega\,\,$---and construct a linearized theory for this case. We also consider the possibility of the Jordan field (scalar field) $\Phi$ to have a nonzero mass. It is well-known that in the massless case of the BD theory in $D=4$ dimensions, the BD parameter $\omega$ must obey the condition $\omega\gtrsim 4\times 10^4$ in order to satisfy the restriction on the PPN parameter $\gamma$ \cite{Skordis}. Such an extremely big value of $\omega$ looks not very natural. It is desirable to generalize the BD theory so that, on the one hand, it satisfies the gravity tests, and, on the other hand, $\omega \sim O(1)$. In the present paper, we demonstrate how to achieve this by means of the extra dimensions.

We show in the massless case that the introduction of $d$ extra dimensions brings new possible values for the BD parameter $\omega$, for which the condition $\gamma=1$ can be fulfilled exactly, as in GR. One possibility is when $\Omega=0$; i.e., the gravity source has a dustlike EoS in all spatial dimensions. In this case, $\omega=-1-1/d$, which, as we show, however corresponds to a ghost scalar $\Phi$. Yet another possibility is the introduction of nonzero EoS parameter $\Omega$ in the extra dimensions. Then, provided that the value of this EoS parameter is fine-tuned to $\Omega = -1/2-1/[2d(\omega+1)]$, we have $\gamma=1$. In this case, it is possible to construct such a solution such that $\Omega>0$, which is impossible in purely metric theories, while the Jordan field $\Phi$ remains nonghost as desired. For this model, moreover, the BD parameter can be of the order of unity: $|\omega| \sim O(1)$. 

In the general massive scalar field case, the metric coefficients acquire the correction terms in the form of the Yukawa potential. The Yukawa mass scale is defined by the mass of the scalar field. Based on the results of the inverse-square-law experiments, assuming that the BD parameter $\omega$ satisfies the naturalness condition $\omega\sim O(1)$, we obtain the lower bound on the Yukawa mass scale, $m\gtrsim 10^{-11}\,$GeV. The experimental constraints on the PPN parameter $\gamma$ require that the EoS parameter $\Omega$ be extremely close to $-1/2$, similar to the usual KK models with Ricci-flat internal spaces \cite{CEZ:2013,CEZ:2014}. We should stress that $\gamma$ (the measure of the space curvature produced by unit rest mass) being equal to unity is not sufficient for the model to satisfy all gravitational tests in the Solar system; see Ref. \cite{Will}. For example, it is also necessary to demonstrate that the PPN parameter $\beta$ (measure of nonlinearity in the superposition law for gravity) is equal or very close to unity. In the present paper, we do not calculate $\beta$ for the model under consideration.

The paper is structured as follows: In Sec. \ref{sec:model}, we specify the generalized BD model and present the background metric coefficients and scalar field. In Sec. \ref{sec:linearized}, we perturb the background model with a pointlike mass and obtain the linearized equations for the metric coefficients and scalar field perturbations. In Sec. \ref{sec:sol}, we get exact solutions of the linearized equations and obtain experimental restrictions on the parameters of the model. The main results are summarized in the concluding Sec. \ref{sec:conc}. 
\section{General setup and background model}
\label{sec:model}
We start with the $D\geq 4$ dimensional gravitational action in the form
\begin{equation}\label{action}
    S_g = \cfrac{1}{2\kappa_D^2}\int {\rm d}^D x \sqrt{|g|}\left[f(\Phi) R+h(\Phi)\,\nabla_M\Phi\nabla^M\Phi-U(\Phi)\right],
\end{equation}
where $g\equiv \det(g_{MN})$, the scalar $\Phi$ couples nonminimally to the scalar curvature $R$; and $f$, $h$, $U$ are some functions of $\Phi$. The constant $\kappa_D$ is defined as $\kappa_D^{2}\equiv 2 S_{D-1} \tilde G_D /c^4$, where $S_{D-1}$ is the total solid angle in the $(D-1)$-dimensional space and $\tilde G_{D}$ is the $D$-dimensional gravitational constant. The total action of the gravitating system is then the sum $S=S_g+S_m$, where $S_{m}=S_{m}[\Psi, g_{MN}]$ is the action of gravitating matter fields $\Psi$.  

For this model, the system of dynamical equations,   
\be{variationMetricFormalism}
\cfrac{\delta S}{\delta g^{MN}}=0\quad\textnormal{and} \quad \cfrac{\delta S}{\delta\Phi}=0,
\ee
takes the following form, correspondingly:
\ba{fieldEqsMetricFormalism}
& f G_{MN}+\left[f''-\cfrac{h}{2}\right]g_{MN}(\nabla \Phi)^2+g_{MN}f'\Delta_D \Phi \nn \\
& +(h-f'')\nabla_M\Phi\nabla_N\Phi \nn \\
&-f'\nabla_M\nabla_N\Phi+\cfrac{1}{2}\,g_{MN}U=\kappa_D^2 T_{MN}, \label{metricEq1}\\
& f' R-h'\,(\nabla\Phi)^2-2h\Delta_D\Phi-U'=0\, , \label{metricEq2}
\ea
where a prime denotes derivative with respect to $\Phi$, $(\nabla \Phi)^2\equiv \nabla_M\Phi\nabla^M\Phi$, $\Delta_D\equiv\nabla_M\nabla^M$ is the $D$-dimensional Laplace-Beltrami operator, $T_{MN}\equiv-(2/\sqrt{|g|})\delta S_m/\delta g^{MN}$ is the energy-momentum tensor (EMT) of matter source, and $G_{MN}$ is the Einstein tensor.

By redefinition of variables, we can always set $f(\Phi)\equiv\Phi$ and $h(\Phi)\equiv -\omega(\Phi)/\Phi$ for some function $\omega (\Phi)$. Then, we continue with the higher-dimensional generalization of the massive Brans-Dicke gravity---namely, with the case $\omega = \, \rm const $, reducing the system of Eqs. \eqref{metricEq1} and \eqref{metricEq2} as follows;
\ba{WildBrans}
    &\Phi G_{MN}+\,g_{MN}\cfrac{\om}{2\Phi}(\nabla \Phi)^2+g_{MN}\Delta_D \Phi
    -\cfrac{\om}{\Phi}\,\nabla_M\Phi\nabla_N\Phi \nn \\
    &-
    \nabla_M\nabla_N\Phi+\cfrac{1}{2}\,g_{MN} U=\kappa_D^2 T_{MN}, \label{fieldEq1} \\
    &R-\cfrac{\om}{\Phi^2}\,(\nabla\Phi)^2+\cfrac{2\om}{\Phi}\Delta_D\Phi
    -U'=0\, . \label{secondFieldEq}
\ea
Performing contraction of Eq. \eqref{fieldEq1} with $g^{MN}$, we obtain
\ba{scalarCurvature}
    \Phi R&=&-\cfrac{2}{D-2}\,\kappa_D^2 T+\cfrac{\om}{\Phi}(\nabla \Phi)^2\nn \\
    &&+\cfrac{2(D-1)}{D-2}\,\Delta_D \Phi+\cfrac{D}{D-2}\, U\, ,
\ea
allowing us to exclude $R$ from Eq. \eqref{secondFieldEq} and then obtain
\begin{equation}
    \left[(D-1)+\om(D-2)\right]\,\Delta_D \Phi= \,\kappa_D^2 T+\cfrac{D-2}{2}\,\Phi U'-\cfrac{D}{2}\,U.  \label{fieldEq2}
\end{equation}

We assume that the spacetime manifold $\mathcal M_D$ is a product manifold $\mathcal M_D=\mathcal M_4\times \mathcal M_d$ with $d\equiv D-4$, and that the background metric on this manifold  has the following factorizable form:
\begin{align}\label{backgroundMetric}
    &\hat g_{MN}dX^M\otimes dX^N= \hat g_{\mu\nu}dx^\mu\otimes dx^\nu+\hat g_{mn}dx^m\otimes dx^n,  \nn \\
    &M, N = 0, ...,  D, \quad \mu, \nu = 0, 1, 2, 3; \nn  \\
    &m, n = 4, ..., D.
\end{align}
Here, $\hat g_{\mu\nu}\equiv \eta_{\mu\nu}=\diag (-1, +1, +1, +1)$ is the Minkowski metric over $\mathcal M_4$, and $\hat g_{mn}$ is the metric over a compact $d$-dimensional Ricci-flat space $\mathcal M_d$:
\be{ricciFlatMetric}
    \hat R_{mn}[\hat g^{(d)}]=0 .
\ee
Hereafter, the hats denote background values. Ricci-flat compactifications encompass a wide class of geometries, including tori and Calabi-Yau manifolds. It is worth recalling that components of the Riemann tensor of the Ricci-flat spaces can be nonzero (e.g., in the case of nonvanishing Weyl tensor \cite{Landau}). 

We assume that the Jordan field (scalar field) of the massive BD theory has a potential of the form
\be{formOfU}
    U(\Phi)=\cfrac{\mu^2}{2}\, (\Phi-\hat \Phi)^2.
\ee
Accordingly, the scalar field $\Phi$ has a mass scale $\mu$, and $\hat \Phi$ defines the position of a stable vacuum of the potential $U(\Phi)$. Hence, $\hat \Phi$ is the background value of $\Phi$. Since the scalar field $\Phi$ determines the strength of the gravitational coupling, its background value $\hat \Phi$ cannot be zero. It can be easily verified that the background metric \eqref{backgroundMetric} together with $\hat \Phi = \text{const.}$ solves the field equations \eqref{fieldEq1} and \eqref{fieldEq2} in the absence of matter---i.e., when $\hat T_{MN}=0$. In order for gravity to be attractive, it is necessary that $\hat \Phi > 0$.

\section{Linearized equations}
\label{sec:linearized}

Now, we consider linear perturbations of the background model. The perturbed metric tensor and scalar field read, correspondingly,
\be{metricTensorPerturb}
    g_{MN} \approx \hat g_{MN} + \delta g_{MN} \equiv \hat g_{MN} + h_{MN}, \,\,\, h^K_L\equiv \hat g ^{KM}h_{ML},
\ee
and 
\be{PhiPerturbed}
    \Phi \approx \hat \Phi + \delta \Phi=\hat\Phi(1+\phi), \quad \phi\equiv\delta\Phi/\hat\Phi.
\ee
We assume that these linear perturbations correspond to the perturbation of the EMT of the gravitating system $\delta T_{MN}$. Since there is no background matter, the energy-momentum tensor coincides with  $T_{MN}=\delta T_{MN}$.

In order to perform linearization, it is convenient to rewrite Eq. \eqref{fieldEq1} by eliminating $R$ from it  with the help of Eq. \eqref{secondFieldEq}:
\ba{fieldEq1modified}
&\Phi R_{MN}
+(1+\om)g_{MN}\Delta_D\Phi
-\cfrac{1}{2}\,\Phi U' g_{MN}-\cfrac{\om}{\Phi}\,\nabla_M\Phi\nabla_N\Phi \nn \\
&-\nabla_M\nabla_N\Phi+\cfrac{1}{2}\,g_{MN} U=\kappa_D^2T_{MN}.
\ea
The linearized field equations \eqref{fieldEq1modified} and \eqref{fieldEq2} take the following forms, correspondingly:
\ba{linearizedFieldEqs}
    &  \delta R_{MN}
    + (1+\om)\hat g_{MN}\hat\Delta_D\phi \nn \\
    & -\cfrac{1}{2}\,\hat\Phi\mu^2\hat g_{MN}\phi- \hat\nabla_M\hat\nabla_N\phi= \cfrac{\kappa_D^2}{\hat\Phi} \delta T_{MN}, \,\, \label{linearFieldEq1}\\
    & \left[(d+3)+\om(d+2)\right]\,\hat\Delta_D \phi  \nn\\
    &\quad\quad\quad = \,\cfrac{\kappa_D^2}{\hat\Phi} \,\delta T_{MN}\hat g^{MN}+\cfrac{d+2}{2}\,\hat\Phi \mu^2\phi.
    \label{linearFieldEq2}
\ea
Here, $\delta R_{MN}$ is the linear perturbation of the Ricci tensor, and it can be written in the form (see Ref. \cite{CEZ:2013})
\ba{perturbedRicci}
    &\delta R_{MN} = -\cfrac{1}{2}\,\hat \Delta_{\mathcal L} h_{MN} +2\,\hat \nabla_{(M} Q_{N)},\\
    & Q_N\equiv\hat\nabla_K h^K_N-\cfrac{1}{2}\, \de_N h^K_K\, ,
\ea
where $A_{(MN)}\equiv(A_{MN}+A_{NM})/2$ and $\hat\Delta_{\mathcal L}$ is the Lichnerowitz operator:
\ba{lichnerowitz}
    \hat\Delta_{\mathcal L} h_{MN}&\equiv& \hat \nabla^K \hat \nabla_K h_{MN} + 2 \hat R_{PMLN}h^{PL}-2\hat R_{P(M}h_{N)}^P \nn \\
    &=& \hat \nabla^K \hat \nabla_K h_{MN} + 2 \hat R_{PMLN}h^{PL}.
\ea
In the last line, we take into account the Ricci-flatness of the background model: $\hat R_{MN}=0$.

In order to eliminate the nonphysical d.o.f. due to the diffeomorphism invariance, we impose the gauge condition
\be{gauge}
    Q_N
     = \cfrac{1}{2}\,\de_N\phi.
\ee
Additionally, without loss of generality, we can set  $\hat \Phi=1$, which is equivalent to the renormalization of constants\footnote{Obviously, since $\hat\Phi>0$ by definition, this renormalization does not alter the signs of $\kappa^2_D$ and $\mu^2$.} $\kappa_D^2/\hat\Phi\to \kappa_D^2$, $\mu^2\hat\Phi\to\mu^2$.
 Then, Eq. \eqref{linearFieldEq1} takes the form
\be{linearFieldEq1Full}
 (1+\om)\hat g_{MN}\hat\Delta_D\phi  -\cfrac{1}{2}\,\hat\Delta_{\mathcal L} h_{MN} 
 -\cfrac{1}{2}\,\mu^2\phi\hat g_{MN}
= \kappa_D^2 \delta T_{MN}.
\ee

Now, we suppose that the matter source of the perturbations of $g_{MN}$ and $\Phi$ is a compact gravitating source, representing an astrophysical object, e.g., the Sun. Since the pressure inside the Sun, $p_0$, is negligible as compared to its energy density, $\varepsilon$, we can assume that the EoS in the external (observable) space is dustlike: $p_0=0$. This is the usual assumption for calculating the PPN parameters $\gamma$ (relevant to the deflection of light and the time delay of radar echoes) and $\beta$ (relevant to the perihelion shift) for orbiting test masses in the Solar System in GR \cite{EZ:2010,Will,Landau}. In the present paper, we calculate only PPN parameter $\gamma$. However, for the sake of generality, we allow that the source may have a nonzero pressure/tension $p_1$ in the internal space, characterized by the corresponding EoS parameter $\Omega=p_1/\varepsilon$. In particular, it was shown in a recent work \cite{CEZ:2013} that in the usual purely metric KK models with Ricci-flat compactification, $\Omega=-1/2$ is the necessary condition for the theory to reproduce the PPN parameter $\gamma=1$ as in the case of GR. Therefore, the EMT of the source is chosen in the following form:
\ba{particleEMT}
    T^M_N = \delta  T^M_N = -\varepsilon  \delta^M_0\delta^0_N +  p_1 \delta^M_l\delta^l_N, \nn \\
    \varepsilon  \equiv \rho c^2 = M c^2\,\cfrac{\delta({\bf r})}{\hat V_d}\,, \quad p_1 = \Omega \varepsilon , \quad \Omega = \textrm{const},
\ea
where $\hat V_d\equiv \int d^d y\sqrt{|\hat g^{(d)}|}$ is the comoving volume element of the (unperturbed) internal space, ${\bf r} = (x^1, x^2, x^3)$ is the position vector in the external space, and $M$ is the mass of the object. Clearly, the corresponding matter source represents a gravitating mass $M$, which is pointlike with respect to the external space and uniformly distributed in the extra dimensions (internal space).

Taking into account the structure of the EMT of the perturbation \eqref{particleEMT}, it turns out that the only nonzero components of the metric perturbations $h_{MN}$ are \cite{CEZ:2013,CEZ:2014b}
\begin{align}\label{metricPerturbStructure}
    & h_{00} \equiv -\hat g_{00} \chi_1 = \chi_1, \quad h_{\tilde\mu\tilde\nu} \equiv \hat g_{\tilde\mu\tilde\nu}\chi_2=\delta_{\tilde\mu\tilde\nu}\chi_2, \nn\\
    & h_{mn}\equiv\hat g_{mn}\chi_3,
\end{align}
(hereafter, $\tilde\mu, \tilde\nu = 1, 2, 3$), with $\chi_{1, 2, 3}$ being some scalar functions of the external space coordinates only. Clearly, since $\delta T_{MN}$ depends only on the coordinates of the external space, these functions, together with $\phi$, also depend only on ${\bf r}$. Obviously, this results in the absence of the Kaluza-Klein massive modes.

Now, taking into account that
\be{reducedRiemann}
    \hat R_{PMLN}h^{PL} = \hat R_{pmln} h^{pl}=\chi_3 \hat R_{pmln} \hat g^{pl}  
   =  \chi_3 \hat R_{mn} = 0
\ee
and then
\be{reducedRiemann2}    
\hat\Delta_{\mathcal L} (f \hat g_{MN})=\hat\nabla^K\hat\nabla_K \hat g_{MN} f
=\hat g_{MN}\hat\Delta_{3}f\, ,
\ee
where $f$ is an arbitrary function of ${\bf r}$, we obtain 
\ba{reducedLichnerowicz}
    &\hat\Delta_{\mathcal L} h_{00}=-\hat g_{00}\Delta \chi_1=\Delta \chi_1, \quad \hat\Delta_{\mathcal L} h_{mn}=\hat g_{mn}\Delta \chi_3,& \nn \\
    &\hat\Delta_{\mathcal L} h_{\tilde\mu\tilde\nu}=\hat g_{\tilde\mu\tilde\nu}\Delta \chi_2=\delta_{\tilde\mu\tilde\nu}\Delta \chi_2 ,\,\,\,\hat\Delta_D\phi=\Delta\phi.&
\ea
Here $\Delta\equiv\hat\Delta_3$ is the Laplace operator over the flat external space. Therefore, the components of the linearized field equations \eqref{linearFieldEq1Full} are
\ba{linearFieldEqsEMT}
    & - (1+\om)\Delta\phi -\cfrac{1}{2}\,\Delta \chi_1 + \cfrac{1}{2}\,\mu^2\phi= \kappa_D^2 \varepsilon ,& \label{linearFieldEq1EMT}\\
    & (1+\om)\Delta\phi - \cfrac{1}{2}\,\Delta \chi_2 -\cfrac{1}{2}\,\mu^2\phi 
    = 0,&\label{linearFieldEq2EMT}\\
    &(1+\om)\Delta\phi-\cfrac{1}{2}\,\Delta \chi_3 -\cfrac{1}{2}\,\mu^2\phi = \kappa_D^2 \Omega \varepsilon .&\label{linearFieldEq3EMT}
\ea
And, for Eq. \eqref{linearFieldEq2}, we have
\ba{linearFieldEq2Full}
    &\left[(d+3)+\om(d+2)\right]\,\Delta \phi-\cfrac{d+2}{2}\,\mu^2\phi \nn \\
    &= -\kappa_D^2\varepsilon(1-\Omega d).&
\ea


\section{Solutions of the linearized equations and experimental constraints}
\label{sec:sol}
The system of four differential equations \rf{linearFieldEq1EMT}-\rf{linearFieldEq2Full} can be solved by reducing it to the system of linear algebraic equations by means of the Fourier transform. All of the four sought solutions can be expressed by one master equation
\ba{solutionStructure}
    f(r) = \cfrac{\kappa^2_D Mc^2}{4\pi \hat V_d}\, \cfrac{1}{r}\left[ A -\left(A-\cfrac{B}{C}\right)e^{-m r}\right],
\ea
where
\ba{constants}
    C \equiv (d+3)+\om(d+2), \quad m^2\equiv\cfrac{d+2}{2}\,\cfrac{\mu^2}{C}\,,
\ea
and the constants $A$ and $B$ are defined for each field separately:
\ba{constantsSolutions}
    \phi(r): & A_0=0, \quad B_0 = 1-\Omega d\,,\label{constantsPhi}\\
    \chi_1(r): &  A_1 = \cfrac{2[1+d(1+\Omega)]}{d+2}, \nn
    \\
            & B_1=2\left[d (\omega +1) (\Omega +1)+\omega +2\right], \label{constantsChi1} \\
    \chi_2(r): & A_2 = \cfrac{2(1-\Omega d)}{d+2}, \nn
    \\
            & B_2=2(1+\omega)(1-\Omega d), \label{constantsChi2} \\
    \chi_3(r): & A_3 = \cfrac{2(1+2\Omega)}{d+2}, \nn
    \\
            & B_3=2\left[ 1+\omega+(3+2\omega)\Omega \right]. \label{constantsChi3}
\ea
Since Eq. \rf{linearFieldEq2Full} is the Helmholtz equation, the scalar field $\phi$ is of the Yukawa potential form with the Yukawa characteristic mass scale $m$. Obviously, to have a physically reasonable solution, we should demand $m^2>0$, leading to the condition
\be{nonGhostCondidion}
C>0 \quad \Rightarrow \quad \omega > \omega_{\mathrm{cr}}=-\cfrac{d+3}{d+2}\,,
\ee
which is exactly the ghost free condition \cite{Capozziello}. Under this condition, the kinetic term of the scalar field yields the ``correct" positive sign in the Einstein frame\footnote{In this study, since we begin with the BD theory, we consider the metric in the Jordan frame---the original frame of the BD theory---as the physical one. In this approach, the scalar field is directly coupled to the scalar curvature in the action. We can decouple it by means of the metric conformal transformation to the Einstein frame. Alternatively, we could start from the very beginning by considering GR with decoupled scalar field and scalar curvature.  Such a multidimensional KK model with a massive, static, spherically symmetric scalar field was investigated in Ref. \cite{zhuk17}, wherein the metric in the original Einstein frame is considered as the physical one. Similar to the present paper, it is found that the massive scalar field contributes to the gravitational potential. However, this correction has the form of the Yukawa potential [similar to Eq. \rf{solutionStructure}] plus some additional terms which are absent in the present paper.}. The admixture of the Yukawa potential to the metric perturbations $\chi_1$, $\chi_2$ and $\chi_3$ is due to the admixture of the scalar field terms to Eqs. \rf{linearFieldEq1EMT}-\rf{linearFieldEq3EMT}. It can be easily verified from Eqs. \rf{solutionStructure} and \rf{constantsChi1}-\rf{constantsChi3} that the combinations $\chi_1+\chi_2$, $\chi_1+\chi_3$ and $\chi_2-\chi_3$ 
behave as $1/r$ without such an admixture, as they satisfy the Poisson equation. Therefore, these combinations define a transition to new variables corresponding to pure gravitational d.o.f. decoupled from the Jordan scalar field that, as known, takes place in the Einstein frame. Here, only the massive scalar field has the form of the Yukawa potential.

Now, we want to investigate under which conditions the above solutions do not contradict the data from observations. It is well known (see, e.g., Ref. \cite{Landau}) that the metric correction term $h_{00}=\chi_1$ defines the gravitational potential: $\chi_1=-2\varphi/c^2$. The inverse-square-law experiments impose restrictions on the Yukawa corrections to the Newtonian gravitational potential \cite{Kapner}. On the other hand, the ratio $\gamma =h_{\tilde\mu\tilde\mu}/h_{00}=\chi_2/\chi_1$ defines the PPN parameter $\gamma$. The gravitational tests, such as the Shapiro time-delay experiment \cite{Bertotti:2003}, tightly restrict this ratio to the value of unity. We make use of the results from these experiments to obtain constraints on the parameters of the model under consideration in this work. To perform it, we consider first the massless case, which is simply the higher-dimensional generalization of the original Brans-Dicke gravity.

\subsubsection{Massless scalar field}
In the case of a massless scalar field, i.e., $\mu=m=0$, the contributions from the Yukawa correction terms disappear and Eq. \eqref{solutionStructure} reads
\ba{solutionStructMassless}
    f(r) = \left(\cfrac{ \kappa^2_D Mc^2}{4\pi\hat V_d}\,\cfrac{B}{C}\,\right) \cfrac{1}{r}.
\ea
Here, the gravitational potential $\varphi = -\chi_1 c^2/2$ has the Newtonian form and should exactly coincide with the Newtonian expression $\varphi_N = -G_N M/r$. To get it, the higher-dimensional $\kappa_D^2$ and the Newtonian $G_N$ gravitational constants are related as
\ba{masslessNewton}
    \cfrac{\kappa_D^2}{\hat V_d}\cfrac{B_1}{C} = \cfrac{8\pi G_N}{c^4}\, .
\ea
Therefore, we cannot make use of the results of the inverse-square-law experiments to obtain restrictions on the parameters of the massless model. Let us then consider the PPN parameter $\gamma$. Since $\chi_2(r) = (B_2/B_1)\chi_1(r)$, for the PPN parameter $\gamma$ we obtain
\ba{masslessGamma}
    \gamma = \cfrac{B_2}{B_1} = \cfrac{(1+\omega)(1-\Omega d)}{d (\omega +1) (\Omega +1)+\omega +2}\, .
\ea
If we set $d=0$ (corresponding to the absence of extra dimensions), the well-known result for the original BD theory in four dimensions,
\ba{BD4D}
    \gamma = \cfrac{\omega+1}{\omega +2},
\ea
is restored as a particular case \cite{Perivolaropoulos}. The precision Shapiro time-delay experiment \cite{Bertotti:2003} restricts the value of $\gamma$ to the narrow interval
\ba{gamma}
    \gamma-1 = (2.1\pm 2.3)\times 10^{-5}
\ea
implying, by means of Eq. \eqref{BD4D}, that $\omega$ must be about $4\times 10^4$ or greater. The values $\omega\sim O(1)$ are, therefore, excluded, which constitutes the problem of naturalness of the original BD theory. On the other hand, we see from Eq. \eqref{masslessGamma} that the presence of extra dimensions may extend the set of allowed values for $\omega$.

First, we consider the case of $\Omega =0$ (the massive source yields no pressure/tension in the extra dimensions). Then the exact equality $\gamma=1$ with $\Omega =0$ from \eqref{masslessGamma} leads to the condition:
\ba{masslessOmega0}
    \omega \equiv \omega_0= -1-\cfrac{1}{d} < -1\,.
\ea
Therefore, in this case the PPN parameter $\gamma$ exactly coincides with the value of GR and, from this point, these theories are indistinguishable. However, parameter $C$,
\be{ghostPhi}
C = (d+3)+\omega_0 (d+2)=-\cfrac{2}{d} <0
\ee
has negative sign, which implies that the scalar field is a ghost. It is worth noting that here $B_1=C$, and their ratio does not change sign in Eq. \eqref{masslessNewton}.

To avoid ghosts, we turn now to the case of nonzero $\Omega \neq 0$. From Eq. \eqref{masslessGamma}, it immediately follows that it is possible to obtain the exact equality $\gamma=1$ provided that the fine-tuning condition
\ba{masslessOmegaNon0}
    \Omega= -\cfrac{1}{2}-\cfrac{1}{2 d (\omega+1)}
\ea
is satisfied. For this value of $\Omega$, we obtain $B_1=B_2=C$. The requirement $\Omega > 0$ leads to a restriction on the allowed values of $\omega$:
\be{OmegaPositive}
    \Omega > 0\quad\Rightarrow\quad\omega_0<\omega < -1\, .
\ee
Parameter $\Omega$ is positive also if $\omega<\omega_0$ along with $\omega>-1$. However, these two inequalities are inconsistent. Since we have $\omega_0<\omega_{\mathrm{cr}}<-1$, we can choose $\omega$ in such a way that
\be{NoGhostNoTension}
    \omega_{\mathrm{cr}}<\omega<-1 \quad \Rightarrow \quad \Omega>\cfrac{1}{d}\, .
\ee
The condition $\omega_{\mathrm{cr}}<\omega$ provides with us the positivity of $C>0$---that is, the absence of ghosts. Hence, in this case, we have both the field $\Phi$ being nonghost, and a positive parameter $\Omega$. The latter condition means that the gravitating source has positive pressure in the internal space rather than tension\footnote{Note that positive values of $\Omega$ are totally excluded in the purely metric models \cite{CEZ:2013, CEZ:2014}, where the condition $\gamma=1$ results in the requirement $\Omega=-1/2<0$.}.

The negative values of $\Omega<0$ result in two types of inequalities for the BD parameter $\omega$:
\be{OmegaNegative1}
\omega<-1,\; \omega<\omega_0 \quad \Rightarrow \quad \omega<\omega_0
\ee
and
\be{OmegaNegative2}
\omega > -1,\; \omega>\omega_0 \quad \Rightarrow \quad \omega>-1\, .
\ee
Obviously, condition \rf{OmegaNegative1} leads to a ghost scalar field; see Eq. \rf{nonGhostCondidion}. Whereas, for $\omega$ from Eq. \rf{OmegaNegative2}, ghosts are absent, and $\omega$ can be positive and of the order of unity: $\omega\sim O(1)$, provided that $\Omega < -1/2$ (tension in the extra dimensions).

\subsubsection{Massive scalar field}
Let us turn now to the general case of massive scalar field. To better understand the structure of the metric coefficients $\chi_1$ and $\chi_2$, we rewrite them in the following form:
\be{chinew}
\chi_i = 
 \cfrac{\kappa^2_D c^2}{4\pi \hat V_d}A_i\, \cfrac{M}{r}\left(1+\alpha_i e^{-m r}\right),\quad i=1,2,
\ee
where
\be{alpha}
\alpha_i \equiv -1+\frac{B_i}{A_i C}
\ee
and
\ba{alpha2}
&-&\alpha_1A_1=\alpha_2A_2 = -\frac{1}{3+2\omega} \nn\\
&+& \frac{d(1+2\Omega)}{d+2}-
\frac{2d(\omega+1)}{(3+2\omega)C}\left[(1+\omega)(1+2\Omega)+\Omega\right]\, .\nn\\
&{}&
\ea
Parameters $A_1$ and $A_2$ can also be expressed as
\be{A}
A_1-1=-\left(A_2-1\right)= \frac{d(1+2\Omega)}{d+2}\, .
\ee

The limit $m\to \infty \Rightarrow \phi\to 0$, and hence corresponds to the GR limit of the model; viz., to the KK model in GR. In this case, the PPN parameter $\gamma = 1$ only if $\Omega=-1/2$ in full agreement with previous works \cite{EZ:2010,CEZ:2013,CEZ:2014}.\footnote{For the model under consideration in this paper, there are two reasons that affect the form of the metric coefficients: (i) the presence of the Jordan field (the scalar field coupled to the scalar curvature), and (ii) the presence of extra dimensions. The variation of the internal space volume (viz., radions) results in the fifth force \cite{Eingorn12}. In the present paper, this corresponds to the nonzero values of the metric coefficient $\chi_3$. Therefore, the absence of contribution from the Jordan field (e.g., in the limit $m \to \infty$) does not guarantee the equality of the PPN parameter $\gamma$ to 1, since this limit results in multidimensional generalization of GR with its inherent problems---e.g., in the form of massless moduli/radions \cite{EZ:2010,Eingorn12}. These problems can be avoided either due to sufficiently large masses of moduli, or with the help of a fine-tuning of parameters of models. It can be seen from Eqs. \rf{constantsChi3} and \eqref{A} that, in the presence of extra dimensions ($d>0$), only if we set $\Omega = -1/2$ will we get $A_1=A_2=1$ and $A_3=0$, which, respectively, lead in the limit $m\to\infty$ to $\gamma=\chi_2/\chi_1=1$ as in the usual GR and $\chi_3=0$.} In the case of a four-dimensional massive BD model---i.e., when $d\to 0$---as follows from Eqs. \rf{alpha2} and \rf{A}, we reproduce the results of Ref. \cite{Perivolaropoulos} [up to evident substitution $\kappa^2_D c^2/(4\pi \hat V_d) \to 2 G_N/c^2$].

Coming back to the general case (i.e., $d\neq 0$ and $m$ is finite), in what follows, we investigate under which conditions the metric coefficients \rf{chinew} do not contradict the gravitational tests relevant to the PPN parameter $\gamma$---the deflection of light and the time delay of radar echoes---in the Solar System. From the inverse-square-law experiments, we can obtain restrictions on the Yukawa correction term \cite{Kapner}. First, at large distances from the gravitating mass, the gravitational potential should have the Newtonian form. Then, keeping in mind that the gravitational potential $\varphi$ is defined by the function $\chi_1: \, \varphi = -\chi_1 c^2/2$, we  define the connection between the higher-dimensional and Newtonian gravitational constants:
\ba{massiveNewton}
\cfrac{ \kappa^2_D }{\hat V_d}\,A_1\,= \cfrac{8\pi G_N}{c^4}\, .
\ea
For this relation to be consistent, we must have $A_1>0$, which is equivalent to $\Omega>-1-1/d$. Moreover, as we shall see below, parameter $\Omega$ should be very close to the value -1/2. For such a value of $\Omega$, we find that $\alpha_1 = 1/C$. Then, if we take the natural value $\omega \sim O(1)$, we get an estimate $\alpha_1 \lesssim 1$ and, consequently, the upper limit on the Yukawa characteristic length of interaction leading to
$\lambda=1/m \lesssim 10^{-3}\,$cm \cite{Kapner}. In other words, the Yukawa mass scale is $m\gtrsim 10^{-11}\,$GeV. This bound is much stronger than the lower limit  $20\times 10^{-27}\,$GeV obtained in Ref. \cite{Perivolaropoulos} only on the basis of the PPN parameter $\gamma$ without taking into account the results of the inverse-square-law experiments.  

It can be easily seen that for the obtained constraint on $\lambda$---i.e., $\lambda\lesssim 10^{-3}\,$cm---the Yukawa correction terms are negligible for the gravitational tests relevant to the PPN parameter $\gamma$ in the Solar System. Indeed, the distance $r$ should be of the order of or greater than the radius of the Sun, $r_{\odot}\sim 7\times 10^{10}\,$cm. Therefore, $r/\lambda \gtrsim 10^{13}$.  Hence, we can drop the Yukawa correction terms in Eq. \rf{chinew} with very high accuracy, and for the PPN parameter $\gamma$ we have
\ba{gammaMassive}
    \gamma= \cfrac{\chi_2(r)}{\chi_1(r)}\approx \cfrac{A_2}{A_1} = \cfrac{1-\Omega d}{1+d(1+\Omega)}\, .
\ea
We note that if $\Omega =0$, we have $\gamma = 1/(d+1)$ that certainly contradicts the observations \cite{EZ:2010, CEZ:2013, CEZ:2014}. Thus, the equality $\gamma=1$ is satisfied provided that  $\Omega = -1/2$. Moreover, supporting this, it was shown in Refs. \cite{EZ:2011,Eingorn10} that a gravitating mass which is pressureless in our space and has the EoS parameter $\Omega=-1/2$ in the internal space does not spoil the stabilization of the internal space.

The negative pressure (viz., the tension) gives rise to a number of questions. This point was discussed in Ref. \cite{Eingorn10}, and afterwards it has been looked for obtaining viable Kaluza-Klein models without such negative pressure in a number of papers (see, e.g., Refs. \cite{EZ:2011,Eingorn:2012ef,CEZ:2014b,Akarsu:2017fjs,Yalcinkaya:2019stk}). However, so far, it is observed that it is very difficult to avoid the EoS parameter $\Omega= -1/2$. Despite the problems about explaining the origin of such an equation of state, the significance of negative pressure (tension) was pointed out in Ref. \cite{Eingorn:2012ip}.

\section{Conclusion}
\label{sec:conc}

In the present paper, we have considered the higher-dimensional generalization of the massive Brans-Dicke theory with Ricci-flat internal space (extra dimensions). The background model was perturbed  by a massive gravitating source which is pressureless in the external (our space) but has an arbitrary EoS parameter $\Omega$ in the internal space. The system of linearized equations for the perturbations of the metric coefficients and scalar field has been solved exactly. Then, the observational data have been considered for obtaining the experimental bounds for the parameters of the model. To this end, we have used the results of both the tabletop inverse-square-law experiments and the experimental limits on the PPN parameter $\gamma$.

First, we have investigated the massless scalar field, which simply generalizes the original four-dimensional Brans-Dicke theory to higher dimensions. In this case, we have shown that the pressureless case, $\Omega=0$, suffers from the presence of a ghost scalar field. On the other hand, we have shown that when $\Omega$ is allowed to be nonzero, it is possible to construct models which, first, are in agreement with the strongest experimental limits on the PPN parameter $\gamma$, and second, have a natural value $|\omega| \sim O(1)$. The price for this is the fine-tuning condition \rf{masslessOmegaNon0}.

Then, we have investigated the general massive scalar field case. It turned out that the metric coefficients acquire the correction terms in the form of the Yukawa potential with a Yukawa mass scale defined by the mass of the scalar field. Based on the results of the inverse-square-law experiments, and assuming that Brans-Dicke parameter $\omega$ satisfies the natural condition $\omega\sim O(1)$, we obtained the lower bound on the Yukawa mass scale leading to $m\gtrsim 10^{-11}\,$GeV. This bound is much stronger than the lower bound  $20\times 10^{-27}\,$GeV obtained in Ref. \cite{Perivolaropoulos} only on the basis of the PPN parameter $\gamma$ without taking into account the results of the inverse-square-law experiments. It is worth noting that this bound we obtained is also valid in the four-dimensional case. The experimental constraints on the PPN parameter $\gamma$ requires that the EoS parameter in the internal space (extra dimensions) $\Omega$ be extremely close to $-1/2$. It is important to emphasize that predicting the gravitational potential in the form of the Newtonian potential and $\gamma \cong 1$ does not suffice to pass the Solar System gravitational tests. Thus, the next step will be to investigate if the PPN parameter $\beta$ could be equal or very close to unity under the conditions we found in this paper for satisfying the PPN parameter $\gamma\cong1$, as suggested by the experiments.

\begin{acknowledgements}
\"{O}.A. acknowledges the support of the Turkish Academy of Sciences in the scheme of the Outstanding Young Scientist Award (T\"{U}BA-GEB\.{I}P). \"{O}.A. also acknowledges the support and hospitality of the Abdus Salam International Center for Theoretical Physics (ICTP), where part of this work was carried out. A.C. acknowledges the support and hospitality of the Istanbul Technical University, where part of this work is carried out. A.C. would also like to thank to A.K. Semenov for useful discussions and valuable comments.
\end{acknowledgements}

\end{document}